\documentclass[aps,floatfix,twocolumn,prb,superscriptaddress]{revtex4-2}

\usepackage{xcolor} 
\usepackage{graphicx}
\usepackage{subfig}
\usepackage{wrapfig}
\usepackage{float}
\usepackage{subcaption}
\usepackage[hidelinks]{hyperref}
\usepackage{amssymb}
\usepackage[utf8]{inputenc}
\usepackage{amsmath}
\newcommand{\abs}[1]{\left|#1 \right|}
\newcommand{\mat}[1]{\mathrm{#1}}

\definecolor{vastkust}{RGB}{0, 48, 80} 
\hypersetup{
  colorlinks,
  citecolor=vastkust,
  linkcolor=vastkust,
  urlcolor=vastkust}

\renewcommand{\vec}[1]{\mathbf{#1}}

\begin{document}

\title{Ultrafast Entropy Production in Non-Equilibrium Magnets}
\author{Finja Tietjen}
\email{finja.tietjen@chalmers.se}
\author{R. Matthias Geilhufe}
\affiliation{Department of Physics, Chalmers University of Technology, 412 96 G\"{o}teborg, Sweden}
\date{\today}
\begin{abstract}
We present an ultrafast thermodynamics framework to model heat generation and entropy production in laser-driven ferromagnetic systems. By establishing a connection between the magnetic field strength of the laser pulse and magnetization dynamics we model time-dependent entropy production rates and deduce the associated heat dissipation in epitaxial and polycrystalline FeNi and CoFeB thin films. Our theoretical predictions are validated by comparison to experimental magnetization dynamics data, shedding light on thermodynamic processes on picosecond timescales. Crucially, we incorporate recently observed inertial spin dynamics, to describe their impact on heat generation in pump-probe experiments. As such, this formalism provides novel insights into controlling heat production in magnetic systems, and contributes to advancing the understanding of non-equilibrium thermodynamics in magnetic systems, with implications for future experimental protocols in spintronics and nanotechnology.
\end{abstract}
\maketitle
\section*{Introduction}

Manipulation and control of magnetic degrees of freedom have been central for nanotechnologies, data storage, information processing, and spintronics~\cite{Walowski2016}. Today, the spin dynamics of magnetic materials can be excited and probed on subpicosecond time-scales~\cite{bossini2023magnetoelectrics,leitenstorfer20232023}, involving ultrafast demagnetization~\cite{Tauchert2022,Wang2020,Kirilyuk2010,Beaurepaire1996}, ultrafast magnetic switching~\cite{Davies2024,Basini2024,Stupakiewicz2021}, and non-linear magnonic effects~\cite{Zheng2023,Schoenfeld2023}. All of these exciting developments are non-equilibrium processes, tightly bound to relaxation and dissipation effects. The latter induce entropy and heat production on the characteristic time-scales of magnetic excitations and motivate the emergent field of ultrafast thermodynamics~\cite{tietjen2024perspective}. 

Entropy production, introduced in the nineteenth century, quantifies irreversibility in thermodynamic cycles, and underpins the Clausius inequality and the second law of thermodynamics. It characterizes macroscopic heat and mass transfer processes~\cite{de2013non}, such as fluid flow and chemical mixing, and also plays a crucial role in information theory~\cite{shannon1948mathematical}. At the microscopic level~\cite{Seifert2012,caprini2019entropy}, stochastic thermodynamics extends these principles by linking entropy production to the random forces that act on individual particles, allowing macroscopic observables to be understood as averages of fluctuating variables~\cite{jarzynski2011equalities,o2022time}. Recently, it has been shown that the paradigm of stochastic thermodynamics can be extended to collective excitations such as phonons, allowing to measure the ultrafast entropy production of phonons in pump-probe experiments~\cite{caprini2024ultrafast}. For magnetic degrees of freedom, stochastic thermodynamics framework has been developed, involving expressions for entropy production and fluctuation theorems~\cite{bandopadhyay2011feedback,bandopadhyay2015macrospin,bandopadhyay2015stochastic,Aron2014}. Furthermore, stochastic thermodynamics of magnets has proven powerful in deriving microscopic spin-dynamics equations involving the interaction of spins with a bosonic bath~\cite{Anders2022}.

\begin{figure}
    \centering
    \includegraphics[height=4cm]{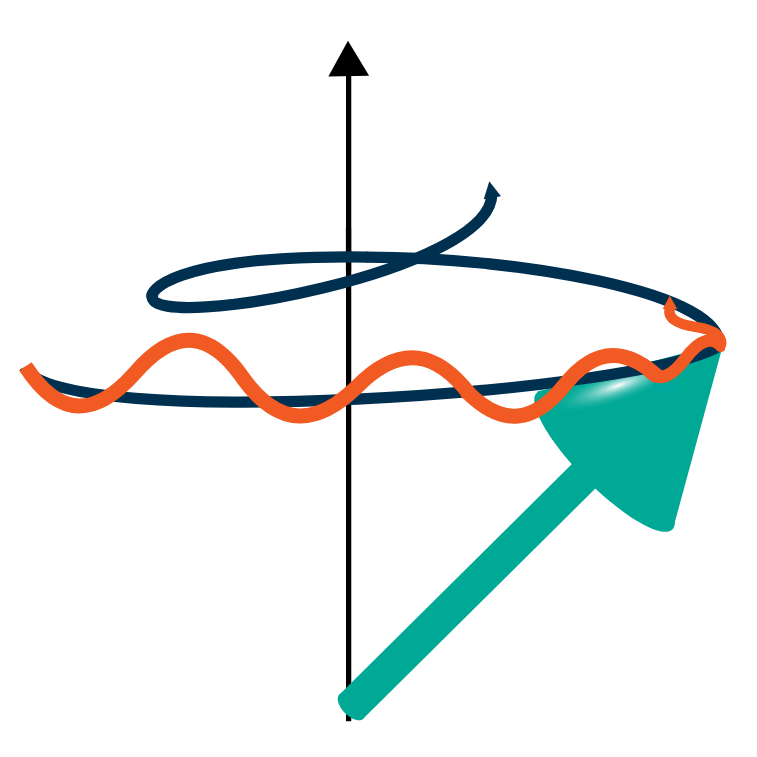}
    \caption{Spin movement including precession (blue) and inertial spin movement (orange) as described by the inertial Landau-Lifshitz-Gilbert equation.}
    \label{fig:nutation}
\end{figure}

Interestingly, while entropy production in the stochastic environment is a well-established concept, measuring entropy production experimentally has remained a challenging task~\cite{manikandan2020inferring, otsubo2022estimating}. Here, we show how entropy production in magnets can be determined in pump-probe experiments, where the magnetization is excited by a pump laser pulse and measured using the magneto-optical Kerr effect (MOKE). To do so, we assume additive noise as proposed by Ma and Dudarev \cite{Ma2012} and develop the corresponding entropy production using a path integral approach~\cite{caprini2019entropy}. Moreover, in contrast to previous work~\cite{bandopadhyay2015macrospin, bandopadhyay2015stochastic, aron2014magnetization}, we include inertial spin movement, i.e., nutation (see \autoref{fig:nutation}), following recent experimental results~\cite{neeraj2021inertial, d2023micromagnetic, li2015inertial}.

Our work is structured as follows. First, we develop the theory for the entropy production in a laser driven magnet including inertial effects. Second, we show how entropy production can be derived from MOKE experiments and compare a theoretical simulation with data taken from Ref. \cite{neeraj2021inertial}. Finally, we discuss the contribution of the inertial effect on the total entropy production.  

\section{Stochastic magnetization dynamics}
The spin dynamics of a material are routinely described in terms of the Landau-Lifshitz-Gilbert equation (LLG)~\cite{gilbert2004phenomenological}. 
Here we consider two extensions: the inertial response and stochastic noise, giving rise to the following equation of motion for the magnetization $\vec{M}$,
\begin{equation}
    \label{eq:inertialLLG}
    \dot{\vec{M}} = - \gamma\,\vec{M} \times \left( \vec{B}  + \eta \dot{\vec{M}} + \eta \tau \ddot{\vec{M}} + \sqrt{2D} \vec{h}\right). 
\end{equation}

The parameters $\gamma$ and $\eta$ are the gyromagnetic ratio and the damping factor, respectively. 
Note that depending on the formulation of the LLG, the damping factor $\eta$ is related to the dimensionless damping factor $\alpha$, by the saturation magnetization $M_s$ and the gyromagnetic ratio $\gamma$, as follows $\eta = \alpha /(\gamma M_s)$.
Magnetic inertial effects describe high-frequency spin-nutation terms~\cite{mondal2017relativistic, faehnle2011generalized, boetcher2012significance, bhattacharjee2012atomistic, ciornei2011magnetization} which have recently been verified in ferromagnetic thin films~\cite{neeraj2021inertial, Unikandanunni2022inertial, li2015inertial, de2024nutation, d2023micromagnetic}. 
Nutation enters the LLG \eqref{eq:inertialLLG} with a term $\sim \tau \vec{M}\times\ddot{\vec{M}}$, involving the angular momentum relaxation time $\tau$ \cite{wegrowe2012magnetization}. 
The inertial spin movement overlaps with precession, while the entire movement is still damped, as depicted in \autoref{fig:nutation}. 
Note that this form of the LLG equation only holds in the case of a harmonic field~\cite{mondal2017relativistic}. 

Damping or dissipation is tightly bound to fluctuations, according to the well-known fluctuation-dissipation theorem. 
Microscopically, the Gilbert damping can be derived from the dissipation of energy from magnetic degrees of freedom to electrons~\cite{Ebert2011} or a bosonic bath, e.g., due to phonons~\cite{Anders2022}. 
This dissipation introduces fluctuations, which are taken into account by adding a stochastic noise term, denoted by $\vec{h}$. 
According to the LLG in equation \eqref{eq:inertialLLG} the magnetic fluctuations only concern the transverse component of the magnetization. 
However, as pointed out by Ma and Dudarev \cite{Ma2012} longitudinal fluctuations can be incorporated, giving rise to the following stochastic Landau-Lifshitz-Gilbert (sLLG), which will be the starting point in our work,
\begin{equation}
    \label{eq:inertialLLG-add}
    \dot{\vec{M}} = - \gamma\,\vec{M} \times \left( \vec{B} + \eta \dot{\vec{M}} + \eta \tau \ddot{\vec{M}}\right) - \gamma M_s \sqrt{2D} \vec{h} \, .
\end{equation}

Furthermore, we consider uncorrelated (white) noise. It obeys the correlation relations
\begin{equation}
\label{eq:noisecorrelation}
\begin{split}
   \langle \vec{h}(t)\rangle &= 0 \\ 
  \left<h_i(t)h_j(t') \right> &= \delta(t-t')\delta_{ij}, \qquad D = \eta k_B T 
  \end{split}
\end{equation} 
with unit $1/\sqrt{s}$ and correlation factor $D$. 
Here, $k_B T$ stems from the thermal nature of the considered noise.

Up to this point, neither the external field nor the internal field in the sLLG are specified, which means that the equation can be adjusted to many different systems and experimental setups.

\section{Entropy production in laser driven magnets}
The explicit form of the sLLG in equation \eqref{eq:inertialLLG-add} allows us to use a path integral approach to compute the entropy production in a laser-driven magnet, similar to the Ornstein–Uhlenbeck process~\cite{caprini2019entropy}. The stochastic form of the differential equation prevents us from obtaining a unique solution. Instead, each realization of the same experiment will give a different path $\underline{\vec{M}} = \left\{\vec{M}\right\}_{t_i}^{t_f}$ in the time interval between the initial time $t_i$ and the final time $t_f$. The probability of observing this path is denoted by $P\left[\underline{\vec{M}}\right]$, assuming that all paths start with a defined initial condition $\vec{M}_0 = \vec{M}(t_i)$. For Gaussian uncorrelated noise, the probability for the trajectory of the noise path is given by
\begin{equation}
  P\left[\underline{\vec{h}}\right] \propto \exp \left( - \frac{1}{2} \int\, \mathrm{d}t\, \vec{h}(t)^2 \right).
\end{equation}
However, the noise path can be expressed in terms of $\vec{M}$ and its derivatives, according to the sLLG \eqref{eq:inertialLLG-add},
\begin{equation}
    \label{eq:noise}
   \vec{h} = - \frac{1}{ M_s \sqrt{2D}}\left[\vec{M} \times \left( \vec{B} + \eta \dot{\vec{M}} + \eta \tau \ddot{\vec{M}}\right) + \frac{1}{\gamma} \dot{\vec{M}} \right]\,,
\end{equation}
and the noise $\vec{h}$ can be formally regarded as a function $\vec{h}\left(\vec{M},\dot{\vec{M}}\right)$. Furthermore, considering all possible noise paths allows us to relate the noise path probability to the probability of the magnetization path \cite{caprini2019entropy, Chaudhuri2016, Spinney2012, spinney2012nonequilibrium}, according to
\begin{equation}
  \log P\left[\underline{h}\right] = \log P\left[\underline{M}\right] + \log \det \frac{\partial \underline{\vec{h}}}{\partial \underline{\vec{M}}}.
\end{equation}
Here, the last term is the Jacobian of the corresponding transformation from $\underline{\vec{h}}$ to $\underline{\vec{M}}$, which can be disregarded in the following~\cite{caprini2019entropy}. Using the paradigm of stochastic thermodynamics \cite{Seifert2012}, the entropy production follows by comparing path probabilities according to
\begin{equation} \label{eq:mep}
    \Sigma =
     k_B \log\left( \frac{P\left[\underline{\vec{M}}\right] }{P_r\left[\underline{\vec{M}}\right]} \right) \, ,
\end{equation}
where $P_r\left[\underline{\vec{M}}\right]$ denotes the probability of realizing the time-reversed path starting at $\vec{M}_f$ and ending in $\vec{M}_0$. As $\vec{M}$ and $\vec{B}$ are time-reversal odd, it follows that $\dot{\vec{M}}$ ($\ddot{\vec{M}}$) is time-reversal even (odd). Hence, we obtain for the forward and backward path probabilities,
\begin{multline}
   P\left[\underline{\vec{M}}\right] \propto  \exp \left( - \frac{1}{4DM_s^2} \int\, \mathrm{d}t\right. \\ \times \left. \left[\vec{M} \times \left( \vec{B} + \eta \dot{\vec{M}} + \eta \tau \ddot{\vec{M}}\right) + \frac{1}{\gamma} \dot{\vec{M}} \right]^2 \right)
\end{multline}
and
\begin{multline}
   P_r\left[\underline{\vec{M}}\right] \propto  \exp \left( - \frac{1}{4DM_s^2} \int\, \mathrm{d}t \right. \\ \times \left. \left[\vec{M} \times \left( \vec{B} - \eta \dot{\vec{M}} + \eta \tau \ddot{\vec{M}}\right) + \frac{1}{\gamma} \dot{\vec{M}} \right]^2 \right).
\end{multline}
Expressing the entropy production in terms of the entropy production rate,
\begin{equation}
    \Sigma = \int \mathrm{d}t\,\dot{\sigma}(t),
\end{equation}
we finally obtain (see App. \ref{app:inertialEP})
\begin{equation}
    \dot{\sigma} = -\frac{1}{T} 
    \left[\frac{\eta\tau}{2} \frac{\partial}{\partial t}\langle\dot{\vec{M}}^2\rangle + \langle\dot{\vec{M}}\cdot \vec{B}\rangle\right].
    \label{entropyproduction}
\end{equation}

The expression \eqref{entropyproduction} is the main theoretical result of our work. 
Here, the entropy production rate resembles the form of heat production \cite{bandopadhyay2015stochastic,caprini2019entropy}, but it has two different origins.
One is the well-known heat produced by the field and the other one is heat produced by the coupling to the medium.
To understand this interpretation better, consider that this type of heat can be understood as the work done on the system. 
As both terms show the same form with different fields, the second derivative also does work on the system.
Since the second derivative stems from the inertial term, which originates in the coupling of system and medium, it can be understood as a form of friction between the studied magnetization and the surrounding medium.

The averages in expression \eqref{entropyproduction} are trajectory averages.
In the integration, we are taking time averages.
Combining those two gives a complete average over the entire space and therefore the average entropy production of the studied system. 
To approximate the first term, i.e., the average over the time-derivative of the magnetization squared, we use the definition of variance $Var(X) = \langle X^2 \rangle - \langle X \rangle^2$. 
Due to white noise, we assume that the variance is constant in time. 
Therefore, it follows that $\frac{\partial}{\partial t}\langle\dot{\vec{M}}^2\rangle = \frac{\partial}{\partial t}\langle\dot{\vec{M}}\rangle^2$ and we can rewrite equation \eqref{entropyproduction} in the more accessible form of
\begin{equation}
    \dot{\sigma} \approx -\frac{1}{T}
    \left[\frac{\eta\tau}{2} \frac{\partial}{\partial t}\langle\dot{\vec{M}}\rangle^2 + \langle\dot{\vec{M}}\cdot \vec{B}\rangle\right]\,.
    \label{entropyproduction2}
\end{equation}
Extending the here-derived formalism from the ferromagnetic case to the anti-ferromagnetic case is possible. 
For this purpose, a second lattice must be introduced, and the two equations must be decoupled, if applicable. 
Based on the two decoupled equations, a similar strategy can be applied.

\section{Measuring entropy production using Magneto-Optical Kerr Effect}
We continue by applying our approach to extract the entropy in pump-probe experiments. More specifically, we consider the experiment by Neeraj \textit{et al}~\cite{neeraj2021inertial}, based on the magneto-optical Kerr effect (MOKE). The corresponding experimental sample geometry is outlined in \autoref{fig:experiment}. 
\begin{figure}[]
    \centering
    \includegraphics[width=0.45\linewidth]{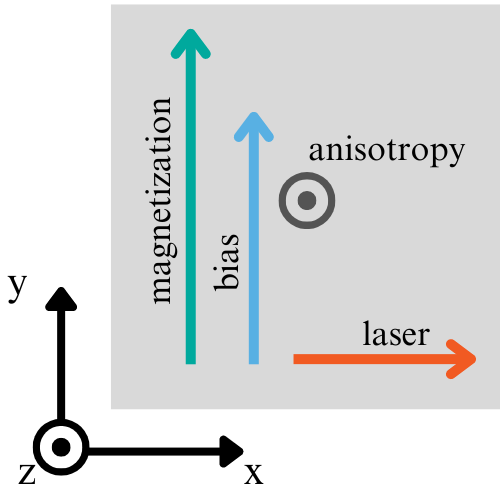}
    \caption{Schematic for experimental setup. The material's thin-film is oriented in the xy-plane with an initial magnetization in y-direction. The thin-film anisotropy is therefore in z-direction. A bias field is applied in the y-direction, while the pump-laser is perpendicular in the x-direction.}
    \label{fig:experiment}
\end{figure}

According to this setup, the magnetic field in equation \eqref{eq:inertialLLG-add} is decomposed in three terms: the effective magnetic field due to the crystalline geometry $\vec{B}_{\text{internal}}$, the bias field $\vec{B}_{\text{bias}}$, and the laser field $\vec{B}_{\text{laser}}$,
\begin{equation}
    \vec{B}(t) = \vec{B}_{\text{internal}} + \vec{B}_{\text{bias}}+ \vec{B}_{\text{laser}}(t) \,.
\end{equation}
We consider a ferromagnetic thin film oriented in the xy-plane. The thin film geometry causes 
a magnetic anisotropy oriented in the z-direction, modeled by
\begin{equation}
    \vec{B}_{\text{internal}} = -D (\vec{M}\cdot \vec{e}_z)\, \vec{e}_z 
\end{equation}
with $D$ being a constant, which we set to $1$ in our simulations.
 The initial magnetization is in the y-direction, caused by a corresponding constant bias field,  
\begin{equation}
    \vec{B}_{\text{bias}} = 0.35 \text{T}\, \vec{e}_y \, .
\end{equation}
In their experiments, Neeraj \textit{et al.}~\cite{neeraj2021inertial} used the TELBE THz laser in the undulator mode producing linearly polarized light with a spectral bandwidth of 20~\%. A typical pulse for a frequency of $\approx 0.8~\text{THz}$ is shown in \autoref{fig:experiment}(a). To compare experiment and theory, we approximate the pulse using a Gaussian envelope with the following functional form,
\begin{equation} \label{eq:fittedlaser}
    \vec{B}_{\text{laser}} = H_0 \sin{(2\pi f t - \phi)} e^{-(t-t_0)^2/(2\delta ^2)}\,\vec{e}_x.
\end{equation}
To approximate the magnetic field strength $H_0$, we assume that the TELBE THz laser operates with a pulse energy of $\lesssim 10~\mu\text{J}$. This pulse energy corresponds to the integrated energy density expressed by $E_{\text{pulse}} = \frac{1}{\mu_0}\int \mathrm{d}V\,\vec{B}_{\text{laser}}^2$, with $\mu_0$ being the vacuum magnetic permeability. Assuming a laser spot diameter of 1~mm, we obtain a magnetic field strength of $B_0 = 0.17~\text{T}$, from comparing \eqref{eq:fittedlaser} with \autoref{fig:experiment}(a). Fitting the other free parameters we obtain: $f = 0.76~\text{THz}$, $\phi= -1.12$, $t_0 = 8$ ps and $\delta = 2$ ps. 
The agreement between theoretical and experimental laser field is shown in \autoref{fig:experiment}(a).

\begin{figure}
    \centering
    \includegraphics[width=0.95\linewidth]{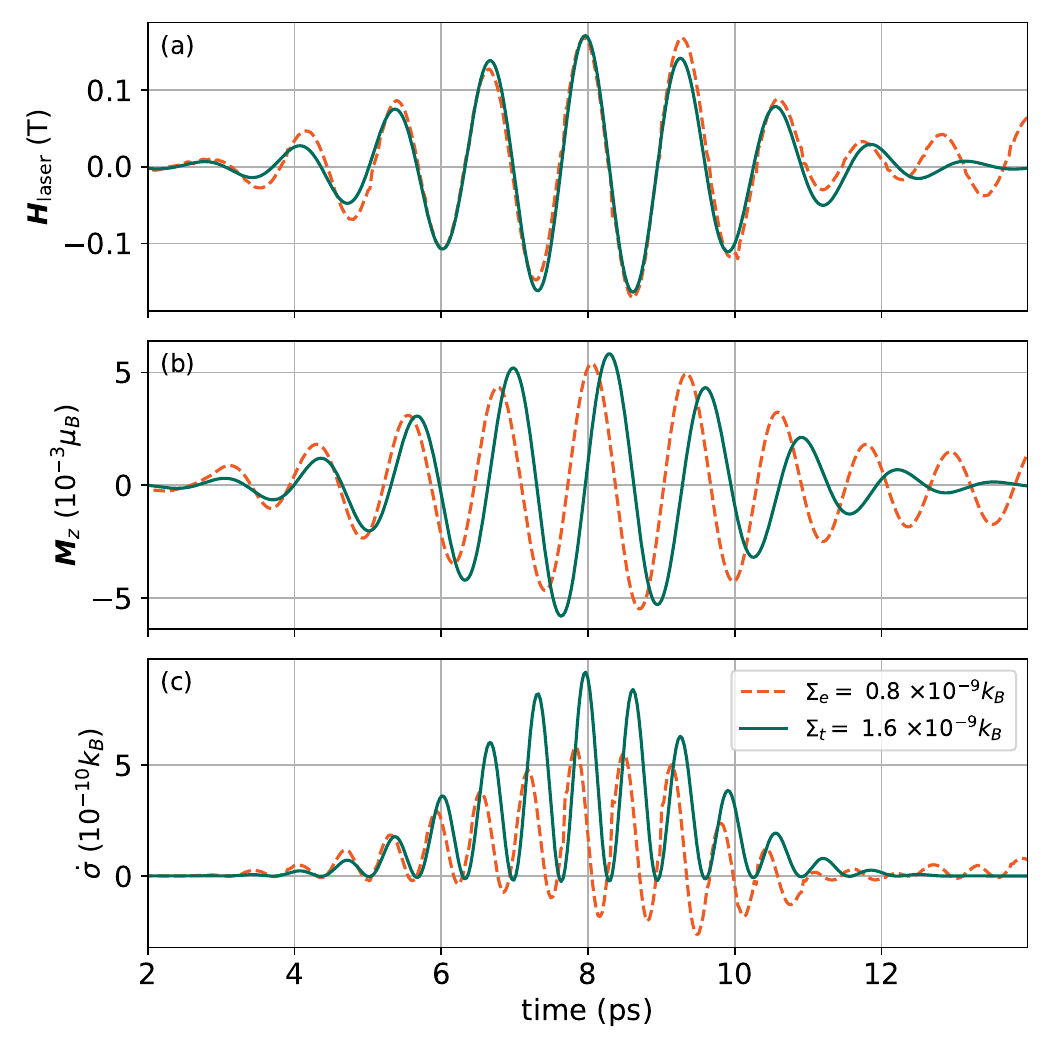}
    \caption{Entropy production per unit cell using magneto-optical Kerr effect. We compare our theoretical model (green solid line) with experimental data \cite{neeraj2021inertial} (orange dashed line). (a) Magnetic field of the THz laser pulse. (b) Reconstructed magnetization. (c) Computed entropy production using equation \eqref{entropyproduction2} for the experimental data and \eqref{eq:entropytheo} for the theoretical estimate.}
    \label{fig:experiment}
\end{figure}

Prior to excitation, the ferromagnetic thin film is magnetized in $y$-direction due to the strong bias magnetic field, $\vec{M} \approx M_s \vec{e}_y$, with $M_s$ the saturation magnetization per unit cell. 
Assuming that the precession term in \eqref{eq:inertialLLG-add} dominates over damping and nutation, we can approximate \eqref{eq:inertialLLG-add} by 
\begin{equation}
\dot{M}_z(t) = \gamma  \vec{B}_{\text{laser}} M_s. 
\end{equation}

Hence, it follows for the entropy production 
\begin{equation}
    \dot{\sigma}(t) = -\frac{1}{T}\left\langle \vec{B}_{\text{laser}} \dot{M_z}(t)\right\rangle = \frac{\gamma M_s}{T}\vec{B}_{\text{laser}}^2(t) \,.
    \label{eq:entropytheo}
\end{equation}

We continue by comparing the theoretical estimate \eqref{eq:entropytheo} for a thin film of polycrystalline NiFe (see \autoref{tab:materials} for parameters), with an entropy production rate \eqref{entropyproduction2} computed from experimental data taken from Neeraj \textit{et al.} \cite{neeraj2021inertial}. 
The latter is obtained from the magnetic field of the TELBE laser (\autoref{fig:experiment}(a)) together with the magnetization determined using MOKE (\autoref{fig:experiment}(b)). 
The result is shown in \autoref{fig:experiment}(c). 
The entropy production rate per unit cell shows oscillations with an overall Gaussian shape. 
As can be seen from the plot, but also from the total produced heat per unit cell, the theoretical result slightly overestimates the experimental data. 
The difference of approximately 50\% arises due to the partially negative entropy production rate obtained for the experimental data. 
However, overall, we can see good qualitative agreement between theory and experiment.

\section{Inertial effects and entropy production}
\begin{figure}[t]
    \centering
    \includegraphics[width=\columnwidth]{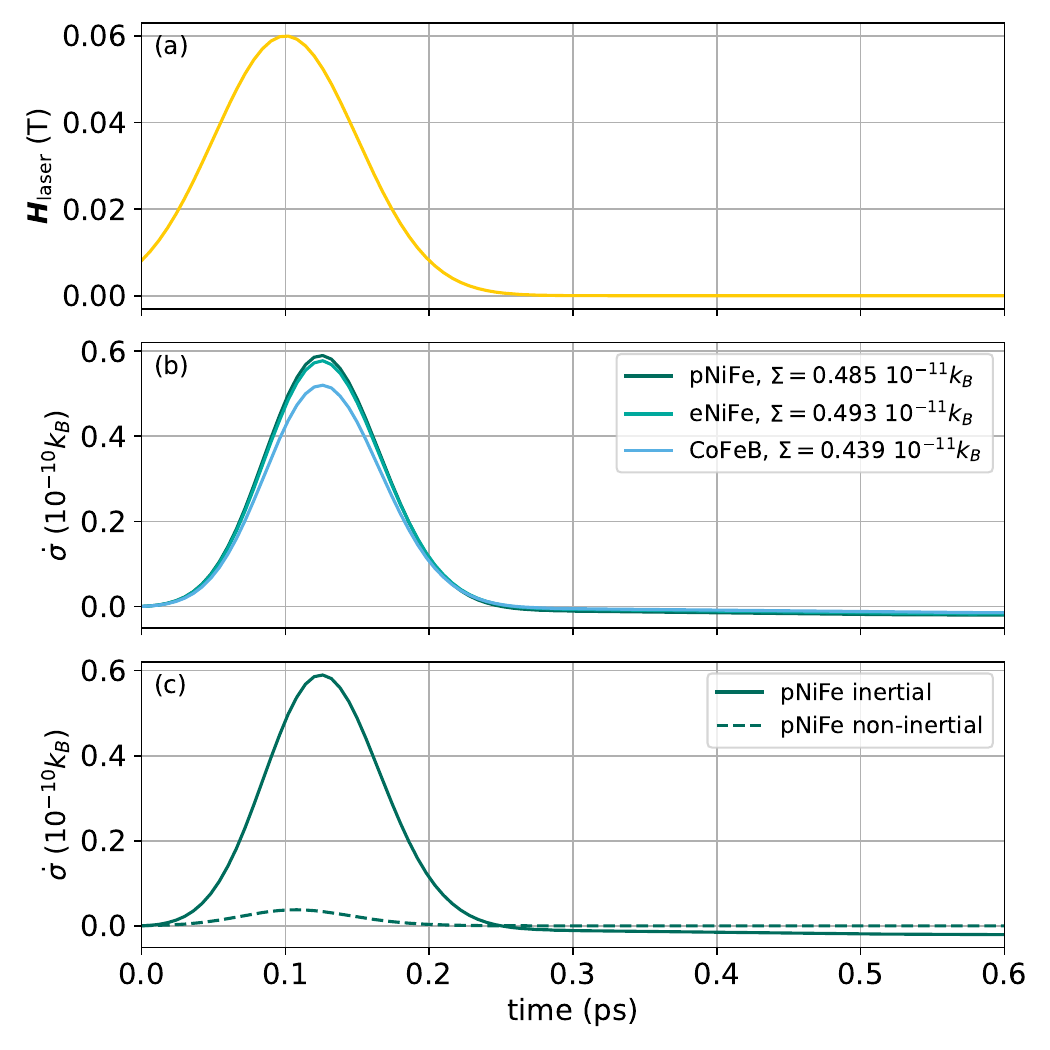}
    \caption{Entropy production per unit cell of nonequilibrium magnets. (a) Magnetic field strength of the pump laser pulse. (b) entropy production rates of polycrytalline NiFe (pNiFe), epitaxial NiFe (eNiFe) and CoFeB. (c) Comparison of entropy production rate for polycrystalline NiFe including ($\tau =12$ ps) and excluding ($\tau=0$ ps) inertial effects. We assume room temperature, i.e., $T=300~K$.}
    \label{fig:EP}
\end{figure}

\begin{table}[]
    \centering
    \begin{tabular}{lcccc}
    \hline\hline 
            & $\alpha$ & $\eta$ ($10^{4}$ Ts/$\mu_B$) & $M_s$ ($\mu_B$) & $\tau$ (ps) \\ \hline
        polycrystalline NiFe & 0.023 & 14.0575& 0.93& 12 \\
        epitaxial NiFe & 0.0058 & 3.545 & 0.93 & 49 \\
        CoFeB & 0.0044  & 2.526& 0.99 & 72 \\
        \hline\hline
    \end{tabular}
    \caption{Simulation parameters, following Ref.~\cite{neeraj2021inertial}. The electron gyromagnetic ratio $\gamma = 2\pi\times 28$ GHz/T is universal and identical for all cases.}
    \label{tab:materials}
\end{table}
\begin{figure}[b]
    \centering
    \includegraphics[width=\columnwidth]{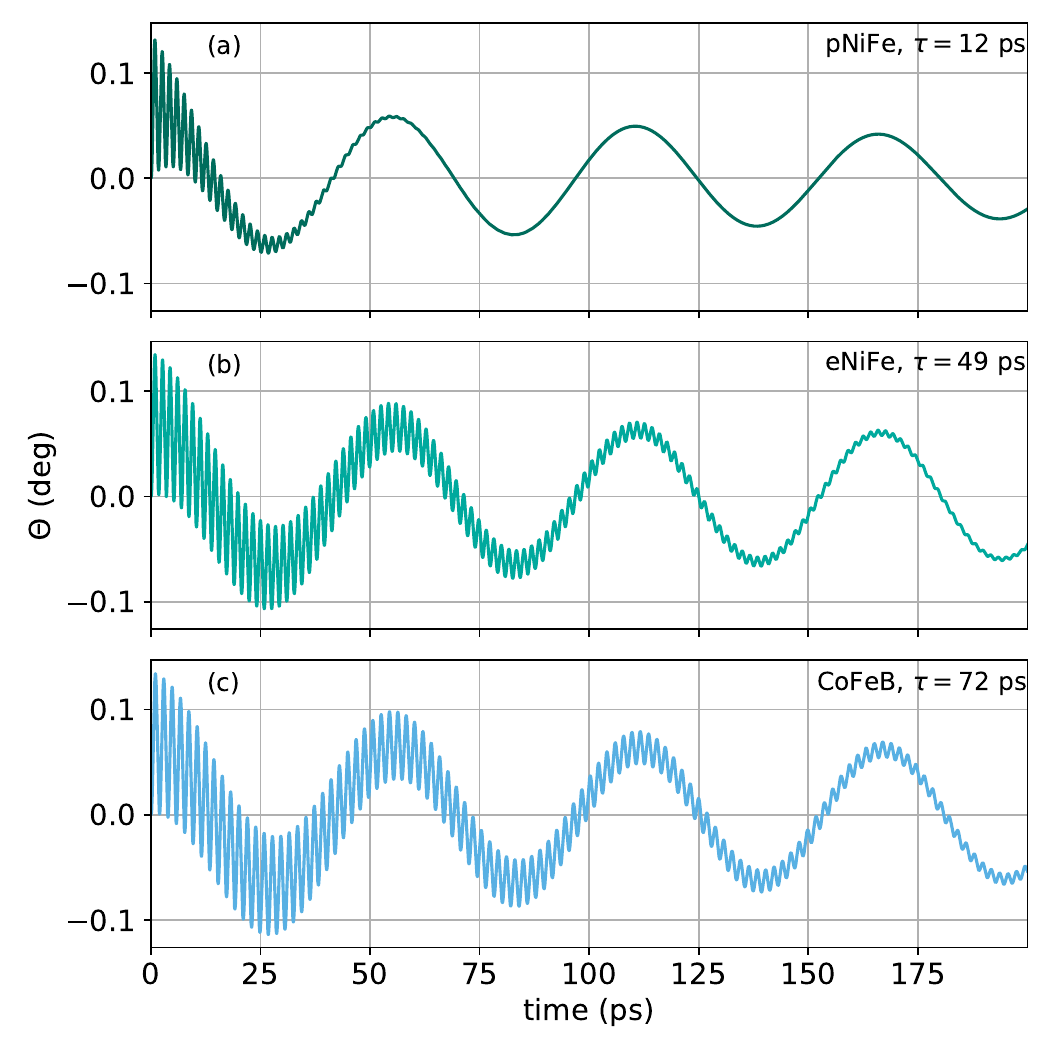}
    \caption{Magnetization dynamics for three different ferromagnetic systems after excitation of a 0.2~ps laser pulse with a magnetic field strength of 0.06~T. The azimuthal angle $\theta = \arcsin \left(M_z/\abs{\vec{M}}\right)$ is plotted. \label{fig:magnetization}}
\end{figure}
To discuss the influence of inertial effects on the entropy production in non-equilibrium magnets, we perform spin-dynamics simulations for three different ferromagnetic systems, summarized in \autoref{tab:materials}: polycrystalline and epitaxial NiFe as well as CoFeB. To do so, we solve the sLLG \eqref{eq:inertialLLG-add} numerically, following the same experimental setup as before (\autoref{fig:experiment}). For the pump laser, we assume a short Gaussian pulse in $x$-direction (\autoref{fig:EP}(a)), 
\begin{equation}
    \vec{B}_{\text{laser}}(t) = H_0 e^{-\frac{(t-t_0)^2}{2 \delta^2} } \,\hat{\vec{e}}_x \, .
    \label{eq:gauss}
\end{equation}
Here, we use $H_0 = 0.06~\text{T}$ and $t_0 = 0.1~\text{ps}$, where the latter is half of the duration of the laser pulse, $\delta=0.2~\text{ps}$.
Note that in the expression for the entropy production, equation \eqref{entropyproduction}, only $\dot{\vec{M}}$ is a stochastic variable. 
Hence, $\left<\dot{\vec{M}}\cdot\vec{B}\right> = \left<\dot{\vec{M}}\right>\cdot\vec{B}$. 
Together with the simplification, $\frac{\partial}{\partial t}\langle\dot{\vec{M}}^2\rangle = \frac{\partial}{\partial t}\langle\dot{\vec{M}}\rangle^2$, leading to equation \eqref{entropyproduction2}, it is sufficient to compute the average of $\dot{\vec{M}}$ which is obtained by neglecting the noise term in equation \eqref{eq:inertialLLG-add}. 
This approach is furthermore justified by the strong bias field leading to small deviations from a fixed equilibrium position.
To obtain a numerically stable solution for the nonlinear second-order differential equation \eqref{eq:inertialLLG-add}, we use the corresponding routines provided in Mathematica 14.

The magnetization dynamics of the three systems are shown in \autoref{fig:magnetization}. The damped oscillation of the magnetization and the influence of the inertial movement are clearly visible. Comparing epitaxial and polycrystalline NiFe, we observe that a higher $\tau$ leads to prominent high-frequency oscillations due to nutation. Note that all three systems get excited to the same angle from the pump-laser. Therefore, all differences in dynamics and consequently entropy production stem from the relaxation behavior. These dynamics have been proven experimentally and previous numerical results agree with our result~\cite{neeraj2021inertial}.
Based on the calculated dynamics, we determine the entropy production rate according to equation \eqref{entropyproduction2}, where we choose a temperature of the crystal (bath) of $T=300~\text{K}$. The results are plotted in \autoref{fig:EP}.

\autoref{fig:EP}(a) shows the Gaussian pump laser pulse with a maximum at $0.1$ ps.
The corresponding entropy production rate per unit cell is shown in \autoref{fig:EP}(b). 
It follows the laser pulse with a delay of $0.026$ ps. 
The maximal entropy production rates are material dependent and decrease with increasing inertial effects $\tau$. 
To investigate this effect, we modeled and compared the entropy production rate of polycrystalline NiFe with and without inertial effects, as shown in \autoref{fig:EP}(c). 
It turns out that inertial effects significantly enhance the entropy production rate. 

To contextualize the entropy production, we relate it to the produced heat in the pump-probe experiment on ferromagnetic thin-films. 
The heat production rate can be computed from the entropy production rate via $\dot{q}=\dot{\sigma}T$, where $T$ is the temperature~\cite{Seifert2012}. 
According to \autoref{fig:EP}(b), the total entropy production in the thin film ferromagnets is $\approx 0.5\times 10^{-11}~k_{\text{B}}$. 
Assuming room temperature $T\approx300$ K, we get a corresponding heat production of $2 \times 10^{-31}~\text{J}$  per unit cell. Consequently, the heat production in the volume enclosed by the $15$ nm thin-film and a laser diameter of $1$ mm is on the order of $1.5 \times 10^{-16}~\text{J}$. We note, that in this setup, the heat production is significantly smaller than the incoming laser energy. For the Gaussian pulse in \eqref{eq:gauss}, we determine for the pulse energy a value of $0.24~\mu\text{J}$. 

\section{Discussion and Conclusion}
We developed an ultrafast thermodynamics framework to model the generated heat and entropy in laser driven ferromagnetic systems. 
We show that these quantities can be determined from the magnetic field strength of the laser pulse and the magnetization dynamics, which can be measured, e.g., using the magneto-optical Kerr effect. 
We modeled the time-dependent entropy production rate and inferred the corresponding heat production for three thin-film ferromagnets, epitaxial and polycrystalline FeNi as well as CoFeB. 
Furthermore, we compared our theoretical results with experimental magnetization dynamics data taken from Ref. \cite{neeraj2021inertial} and used these data to estimate the entropy production in these experiments. 
As a result, our formalism sheds light on thermodynamic processes on the picosecond time scale. 

Our theory incorporates recently detected inertial effects in magnetic systems and describes their influence on heat production in pump-probe experiments. 
Hence, we imagine that this formalism can be guiding for future experiments, either by determining protocols to avoid sample heating or for tuning heat production, e.g., in connection to demagnetization experiments. 

Our theory has been developed under the assumptions of additive (transversal and longitudinal) noise and uncorrelated noise. 
An extension to multiplicative noise could be achieved, e.g., by merging our formalism with the work by Bandopadhyay \textit{ et al.} \cite{bandopadhyay2015macrospin,bandopadhyay2015stochastic}. 
Furthermore, the assumption of white noise requires the coupling of the magnet to a phononic bath with an approximately constant density of states over a large frequency range. 
Considering e.g. magnon-phonon coupling, this assumption has been challenged~\cite{Anders2022}. 
However, following Ref. \cite{caprini2019entropy}, we argue that our formalism of entropy production in driven magnets can be extended to the case of correlated noise by assuming a noise correlation function $v_{ij}(t-t') = \left\langle h_i(t) h_j(t') \right\rangle$ together with its inverse $v^{-1}(t-t')$, $\sum_j \int \mathrm{d}s\, v^{-1}_{ij}(t-s) v_{jk}(s-t') = \delta(t-t')\delta_{ik}$. 
The corresponding entropy production rate given in equation \eqref{entropyproduction2}, will take the form of a convolution, $\dot{\sigma} = \dot{\sigma}_\tau+\dot{\sigma}_B$, with $\dot{\sigma}_\tau \sim \dot{\vec{M}}(t)\left(\mat{v}\ast\ddot{\vec{M}}\right)(t) + \left(\dot{\vec{M}}\ast\mat{v}\right)(t)\ddot{\vec{M}}(t)$ and $\dot{\sigma}_B \sim \dot{\vec{M}}(t)\left(\mat{v}\ast\vec{B}\right)(t) + \left(\vec{B}\ast\mat{v}\right)(t)\dot{\vec{M}}(t)$.

\begin{acknowledgments}
We thank Lorenzo Caprini, Juliette Monsel and Stefano Bonetti for fruitful discussions and insights.

We acknowledge support from the Swedish Research Council (VR starting Grant No. 2022-03350), the Olle Engkvist Foundation (Grant No. 229-0443), the Royal Physiographic Society in Lund (Horisont), the Knut and Alice Wallenberg Foundation (Grant No. 2023.0087), the Carl Trygger Foundation (CTS 23:2462) and Chalmers University of Technology, via the department of physics and the Areas of Advance Nano and Materials. 
\end{acknowledgments}

\appendix

\section{Entropy production for inertial system} \label{app:inertialEP}
Using equation \eqref{eq:mep}, the resulting entropy production rate is
\begin{align}
   \label{eq:EP-inertial}
   \Dot{\Sigma} &= \frac{k_B}{M_s^2 D} \eta \left[ (\vec{M}\times \vec{B})\cdot(\vec{M}\times\dot{\vec{M}})+ \right.\\ \left. \right. \nonumber & \left. \eta\tau (\vec{M}\times\dot{\vec{M}}) \cdot (\vec{M}\times \ddot{\vec{M}}) +
 \frac{1}{\gamma}(\vec{M}\times\dot{\vec{M}}) \cdot \dot{\vec{M}}\right]\, \\ \nonumber
   &= -\frac{1}{T} \left[\eta\tau (\dot{\vec{M}} \cdot \ddot{\vec{M}}) + (\dot{\vec{M}}\cdot \vec{B})\right]\\
   &= -\frac{1}{T} \left[\frac{\eta\tau}{2} \frac{\partial}{\partial t}\dot{\vec{M}}^2 + (\dot{\vec{M}}\cdot \vec{B})\right],
\end{align}
where we used that $(\vec{a} \times \vec{b}) \cdot(\vec{c} \times \vec{d}) = (\vec{a} \cdot \vec{c})\cdot(\vec{b}\cdot\vec{d})-(\vec{a}\cdot\vec{d})\cdot(\vec{b}\cdot\vec{c})$. 
\begin{table}[b]
    \centering
    \begin{tabular}{c|c}
         &  damping \\ \hline
      Neeraj et al. \cite{neeraj2021inertial}  & $\eta = \frac{\alpha}{\gamma M_s}$  \\
      Bandyopadhyay et al. \cite{bandopadhyay2015stochastic} & $\eta' = \frac{\alpha}{M_s}$, $\eta = \alpha M_s$\\
      Gilbert \cite{gilbert2004phenomenological} & $\eta = \frac{\alpha}{\gamma M_s}$\\
      this paper & $\eta = \frac{\alpha}{\gamma M_s}$, $\eta' = \gamma \eta = \frac{\alpha}{M_s}$
    \end{tabular}
    \caption{Conversion between different conventions for damping parameter in sLLG equation.}
    \label{tab:conversions}
\end{table}

\section{Units of Entropy Production Rate}
The units of all quantities used are
\begin{align}
    [\vec{B}] &= \text{T} \, , \quad [M] = \frac{\text{eV}}{\text{T}}\, , \quad [D] = \text{T}^2\text{s} \\ \nonumber
    [\vec{h}] &= \frac{1}{\sqrt{\text{s}}}\, \quad
    [\gamma] = \frac{1}{\text{Ts}}\, , \quad 
    [\eta] = \frac{\text{T}^2 \text{s} }{\text{eV}}\, , \quad 
    [\tau] = \text{s} \, ,
\end{align}
where the magnetization is considered per unit volume.
These units agree with the expectation based on the involved physical quantities and other (experimental) work \cite{neeraj2021inertial}.
The entropy production rate $\dot{\sigma}$ is in units of $k_B$/s per unit cell volume.

Note that the different formulations of the stochastic LLG equation entail different values for the gyromagnetic ratio and the damping factor.
Considering the formulations of Neeraj et al. \cite{neeraj2021inertial}, Bandopadhyay et al.\cite{bandopadhyay2015stochastic}, and the original formulation by Gilbert \cite{gilbert2004phenomenological}, we arrive at the conversions displayed in \autoref{tab:conversions}.

\end{document}